# Influence of the organic cation disorder on photoconductivity in ethylenediammonium lead iodide, $NH_3CH_2CH_2NH_3PbI_4$

Anastasiia Glushkova,[a] Alla Arakcheeva,[*a] Philip Pattison,[bc] Márton Kollár,[a] Pavao Andričević,[a] Bálint Náfrádi,[a] László Forró[a] and Endre Horváth[a]

a EPFL SB IPHYS LPMC, Lausanne, CH-1015, Switzerland. E-mail: alla.arakcheeva@epfl.ch

b SNBL, ESRF, 71 Avenue des Martyrs, Cedex 9, Grenoble, 38043, France

c EPFL SB IPHYS LQM, Lausanne, CH-1015, Switzerland

Corresponding Author:

*E-mail: alla.arakcheeva@epfl.ch

**ABSTRACT**

We report the synthesis and crystal structure of an organic–inorganic compound, ethylenediammonium lead iodide, NH3CH2CH2NH3PbI4. Synchrotron-based single crystal X-ray diffraction experiments revealed that the pristine and thermally treated crystals differ in the organic cation behaviour, which is characterized by a partial disorder in the thermally treated crystal. Based on current–voltage measurements, increased disorder of the organic cation is associated with enhanced photoconductivity. This compound could be a potential candidate for interface engineering in lead halide perovskite-based optoelectronic devices.

**INTRODUCTION**

Widespread attention has been devoted recently to semiconducting organic–inorganic lead halide perovskites, because of the excellent light absorption properties of these materials in the visible and near-infrared regions.1,2 An increasing number of investigations target the precise mapping of the band structure of single crystals. In parallel, the tunability of the crystal morphology and optical bandgap of this class of material is appealing for integrating them as the active components of various optoelectronic devices. Lab-scale prototypes of perovskite solar cells achieved remarkable power conversion efficiencies of 22.7%.3 Recently, perovskite-based photodetectors were found to detect gamma-ray, X-ray, UV, visible and the NIR region with responsivity of 242 A W−1 at low bias (−1 V).4,5 Unlike the symmetrical contact devices, the micro fabricated hybrid devices of methylammonium lead iodide nanowires grown on individual carbon nanotubes and CVD graphene showed extremely high 2.6 × 106 A W−1 responsivities in the visible range, holding the potential to accomplish single photon detection in the future.6–8

The chemical formula of the so called three-dimensional (3D) organic–inorganic halide perovskites is ABX3, where B is a metal (typically Pb2+ or Sn2+) and X are typically halogens (Cl− , Br− or I− ). The A position is occupied by an organic amine cation. Larger organic cations such as the ethylenediammonium cannot incorporate into a 3D perovskite rather they will induce the formation of two- and onedimensional structures in which PbI6 octahedra form layers and columns, respectively.9,10 Recently, blended samples of 3D perovskites and lower dimensional counterparts showed enhanced photostability and environmental stability of perovskite solar cells.11–13 However, the precise phase composition of these blends, namely, their lower dimensional perovskite components is still unknown. Hence, synthesis and structure investigation of the lower dimension perovskites are topics of current interest.

It has been shown that the inorganic part of the structures mainly determines the charge mobility, band gap tunability and the optical and magnetic properties.14 The organic cation determines the luminescence and the electronic properties of the material.15,16 For instance, the choice of the organic cation might influence the dimension of the inorganic part and interactions between the inorganic and organic constituents.15,17 The most important type of these interactions is the weak hydrogen bond. As a consequence of the weak bonds, the organic cation is relatively mobile in the crystal structures and its orientation might also depend on external stimuli.18 It appears that the rotation of the cation in 3D perovskites changes the bandgap from direct to indirect affecting the lifetime of charge carriers (i.e. diffusion and absorption lengths). Duan et al. reported that the orientation disorder of the organic cation strongly influences the dielectric and emission features.19 These reports suggest that the photovoltaic characteristics can also be affected by the organic cation disorder. The cation disorder varies with the preparation method and post treatment of the compound. The cation (or in-cation) disorder is directly related to the atomic displacement parameters (ADP), which can be found from a precise crystal structure determination. Hence, the combination of the photovoltaic characteristic measurements with a precise crystal structure determination of two compounds differing by only the state of the organic cation is a challenging study, which we present here.

We report the crystal structure, (NH3CH2CH2NH3)PbI4 ethylenediammonium lead iodide (hereinafter EDPbI4), in its pristine and thermally treated (1 hour at 423 K followed by cooling) state. We show that the thermal treatment results in stabilization of additional N–H⋯I hydrogen bonds. These bonds facilitate both the splitting of the ethylenediammonium (ED) cation and increased mobility of the carbon atoms. In turn, the current density–voltage characteristics (J–V curves) of the both pristine and thermally treated crystals mean that the increase of the ED leads to an improvement of the photoconductivity in the thermally treated crystal.

**EXPERIMENTAL**

**Synthesis and crystalization**.

Ethylenediammonium lead tetraiodide NH3CH2CH2NH3PbI4 (EDPbI4) crystals were synthesized by solution growth method, 3.3 mmol leadIJIII) acetate trihydrate (PbIJac)2-x-3H2O, >99.9%) was reacted with a 6 ml saturated HI solution (57 wt% HI in H2O). The ice-cooled solution of PbI2 (278 K) was reacted with the respective amount of the ethylenediamine solution (3.30 mmol). At this step, the cooling of the solution is important to avoid the evaporation of ethylenediamine during the rapid exothermic reaction. Small, submillimetre-sized crystals were immediately precipitated and recrystallized to larger millimetre and centimetre sized crystals at 323 K for 7 days. After harvesting, the crystals were wiped with laboratory wipes and dried at room temperature. Two distinct types of crystals were observed (Fig. 1a). At the colder part of the vial (Fig. 1b) flake-shaped, translucent colour crystals were grown. At the warmer part of the solution (Fig. 1c) large bright yellow colour cube-shaped crystals were grown. The powder X-ray diffraction (PXRD) patterns (Fig. 1d and e and more detailed in Fig. S1†) confirm their difference. In this paper, we present detailed analysis of only the cube-shaped crystals. Some characteristics of the flakeshaped phase including PXRD, chemical composition and thermogravimetric analysis are given in ESI.† The cubeshaped crystals were annealed at 323 K for one hour in air resulting in the pristine EDPbI4 crystals; a part of the crystals were further heated at 423 K for one hour in air to produce thermally treated crystals.

**Chemical composition**

The X-ray fluorescence (XRF) spectrometer (EDAX Orbis Micro X-ray Fluorescence Analyzer) was used for determination of Pb and I in the analysed crystals. The energy dispersive X-ray (EDX) spectra (Fig. S2†) revealed the ratio Pb : I = 1 : 4 confirming of $PbI_4$ in the compound. Confirmation of the ED cation presence has been done by the single crystal structure determination.

**Thermogravimetry**

The TGA analyses were done on a Perkin Elmer TGA 4000 system with an auto sampler in $N_2$ atmosphere. The temperature program was from 313 K to 573 K with a heating rate of 10 K min−1 , and the sample was kept at 573 K for 20 minutes. The TGA curves of the titled compound are shown in ESI† (Fig. S3a).

**Single crystal X-ray diffraction**

The synchrotron radiation single crystal X-ray diffraction experiments were carried out at room temperature with wavelength λ = 0.7129 Å using the PILATUS@SNBL detector at the Swiss-Norwegian Beam Lines, European Synchrotron Radiation Facility.20 CrysAlisPro program package was used for the experimental data processing.21 Structural calculations were made with JANA2006 software.22 The main details on the experimental data and crystallographic characteristics of pristine and thermally treated $EDPbI_4$ crystals are given in Table 1. The further experimental details are available as a CIF file by CSD-numbers 434179 and 434178.

**Structure solution and refinement**

The structure has been solved using the Superflip program.23 Arrangement of the $PbI_6$-octahedra (Fig. 2) was found in the space group R$\bar{3}$c, identical for the pristine and thermally treated crystals. However, 57 and 63 independent reflections with I > 3σ(I) violate the c-glide plane for the pristine and thermally treated crystals, respectively. Correspondingly, the space groups R32, R$\bar{3}$ and R3 have been tested. The best results of the structure determination and refinement have been obtained in R$\bar{3}$ for both crystals. Further lowering of the symmetry to R3 decreases the reliability index R1 from 3.14 and 3.98 down to 2.9 and 3.7%, respectively, for the pristine and thermally treated crystal. However, this decrease is associated with the essential increase of a number of refined parameters and does not influence the main observations, which concern the atomic displacements in the ED cation. Hence, we use the R$\bar{3}$ space group for both the pristine and thermally treated crystals. Positions of N and C atoms were found from the difference electron density calculated after localization of Pb and I atoms (Fig. 3a). Refinement of Pb, I, N and C atoms reveals unusually elongated ADP ellipsoids for N and C in the structure of the thermally treated crystal (Fig. 2 and dynamic model in Fig. 3b). We examine the split cation model (i.e. static disorder of the cation), when two its positions of 0.5 occupancy for each of them are refined along with atomic parameters of all atoms, N1, N2, C1and C2, constituted the cation (split cation in Fig. 3b). We also consider a model with splitting of only for C-positions owing to small splitting of N-atoms (0.2 Å) in the split cation model (split C1 & C2 in Fig. 3b). The residual electron density is indifferently low in all three models of the cation disordering in the thermally treated crystal (Fig. 3c). Positions of hydrogen atoms were found in the 1.2 Å vicinity of the carbon and nitrogen atoms and then constructed according to expected geometry of the ethylenediammonium cation, $NH_3CH_2CH_2NH_3$. The hydrogen positions were restricted to 0.96 Å distance and tetrahedron configuration relating to the corresponding carbon and nitrogen atoms taken into account torsion angles. Details of the structure refinements appear very similar for all three models. In Table 1, we present only split cation model as preferable based on consideration of the cation geometry. Geometry characteristics of the ethylenediammonium cation are listed in Table 2 for the pristine crystal and three models of the thermally treated crystal. Characteristics of atomic positions and interatomic geometry are available from the CIF files and ESI† (Tables S1 and S2). The residual electron density calculated after both crystal refinements shows additional weak maxima of about 1 e Å−3 . These maxima were associated with the presence of a small amount of water. Placing the corresponding oxygen position gives about 13(2)% occupation for both crystals. This occupation corresponds to 0.043(4) and 0.043(5) H2O per a chemical formula of the pristine crystal and thermally treated crystal, respectively (Table 1).

**Powder X-ray diffraction**

The PXRD data was collected on an Empyrean system, equipped with a PIXcel-1D detector (Cu Kα-radiation, Bragg– Brentano beam optics) for the pristine EDPbI4 crystals at room temperature. The XRD profile has been successfully fitted by the Rietveld refinement of the model obtained from the single crystal structure determination (Fig. S1†). The whole set of atomic parameters has been refined again for the pristine crystal. No significant difference was found within two standard deviations. This confirms the structure and composition of the studied crystalline phase.

**Photocurrent measurements**

In order to characterize the photoconductivity response of the EDPbI4 single crystals, two-point electrical resistivity measurements (current density vs. voltage, J–V) were performed in the dark and under UV light illumination. All measurements were performed in ambient conditions at room temperature. Golden wires connected with Dupont® 4929 silver epoxy were used as electrical leads. A Keithley 2400 source meter allowed to measure the current with < 0.1 nA resolution, while tuning the applied bias voltage. I–V photocurrent characteristics were obtained under illumination from a UV light source (365 nm) with an intensity of 0.055 mW mm−2 . The voltage was swept from 0 V to +10 V/−10 V and back with a scanning speed of 1 V s−1 . All measurements were performed with crystalline samples of approximately equivalent size. Recalculation from current flow to current density was done by measuring the geometric factor of the sample and considering the absorption depth of the photon with the energy 4 eV in the material.24

**RESULTS AND DISCUSSION**

**Crystal structure of ethylenediammonium lead iodide**

The crystal structure of the cube-shaped EDPbI4 is topologically identical for the pristine and thermally treated crystals and is shown in Fig. 2. The PbI6-octahedra share a common face to form three-octahedra segments of a chain along the hexagonal c-axis. The segments are separated from each other by a distance of about 8.2 Å between the Pb atoms along the chain and about 8.4 Å between their axes. The trigonal prismatic cavity separating two segments of octahedra is not completely empty: a small amount of H2O (about 13% occupation in both crystals) was found in the three-fold axis closer to the face of one of the octahedra (O–I ≈ 3.2 and 3.7 Å). The ED cations are distributed between the chains (Fig. 2). Each organic cation is connected by N–H⋯I hydrogen bonds to I− anions forming PbI6 octahedra. These H-bonds are different in pristine and thermally treated crystals. In the thermally treated crystal, they can be recognized also differently depending on model of the atomic ordering in the ED cation (Fig. 4 and Table S2†). Selection of the correct model needs a special discussion, which is following below.

**Disorder of the ethylenediammonium cation in the thermally treated crystal**

Criteria of the quality of the structure refinement cannot help to select the preferable model of the ED cation disorder among three of them described as dynamic, split cation and split C atoms (Fig. 3).

Each of the models is characterized by the identically low R-factors of 0.04 and reasonably low residual electron density of −2.5 < Δρ < 1.4, which is located at about 1 Å from Pb atoms. Therefore, we consider the geometry of the ED cation (Table 2) and atomic displacement parameters of C and N atoms constituting this structural unit (Fig. 5). Too short distance C1–C2 = 1.29 Å, too large angles N–C–C of about 120° instead of expected 108° (Table 2) and flattened ADP ellipsoids of both C1 and C2 (Fig. 5) define that the dynamic model with thermal atomic vibrations is not suitable. On other hand, the minimum deviation of the N and C atoms from their common plane (0.03(5) Å, Table 2) appears exactly in the dynamic model of disorder. It means the improved geometry of ED cation costs deviations from the plane. This is exactly observed in both the split cation and split C atoms models. They are characterised by the essentially better geometry of the cation (N–C = 1.4–1.6, C–C = 1.4–1.5 Å, angles N–C–C = 108–117°, Table 2) with the cost of 0.15(5)–0.18(3) Å deviation of C2 from the N1N2C1 plane. The deviation of C atoms from the ED plane can be regarded as the reason for their "pedal-like" (ref. 25) oscillations around positions inside the plane. Note that the deviation of 0.13(3) Å from the plane is also observed for C2 in the pristine crystal structure, which is characterized by a good geometry of the ED cation (Table 2). This can be a reason for the stabilization of the thermally treated crystal under the ambient conditions: heating creates the oscillation in the unstable state of the pristine crystal. It can be seen that in both its split models, the virtually undistorted ADP ellipsoid of C1 (Fig. 4 and 5) can be adopted in both static and dynamic types. However, the strongly enlarged ellipsoid of C2 (Fig. 4 and 5) unambiguously indicates at least the contribution of the dynamic type. Considering possible N–H⋯I interactions in two split models (Fig. 4), one can say that two split positions of ED cation are perfectly bonded to the same I atoms. This means that both static and dynamic disorder types of ED cation can be stabilized by the very similar N–H⋯I hydrogen bond interactions. Taken into account the more reasonable distances and angles within ED cation (Table 2), we recognize the split cation model as the best approximation. However, some contribution of the "pedal-like" oscillation of carbon atoms cannot be completely excluded.[25]

Summarizing this analysis, one can conclude that despite the selection of the atomic disorder model in ED cation, the ED cation is certainly disordered in the thermally treated crystal. It is important to note that, in average, the $U_{eq}$ atomic displacement parameters of C and N atoms are systematically larger in comparison to the pristine crystal (Fig. 5). Thus, disorder of ED increases in the thermally treated crystal.

**Correlation between the organic cation disorder and photoconductivity**

The J–V curves (Fig. 6a) show that the photo-induced current is more than one order of magnitude weaker than in the commonly used $CH_3NH_3PbX_3$ material (Fig. S5† shows J–V for $CH_3NH_3PbCl_3$, synthesis of the crystal is described by Kollár et al.[26]) for both the pristine and thermally treated crystals. However, despite the identical amount of water in crystals (Table 1), the thermally treated crystal shows higher photoconductivity. As shown above, the main difference between the crystals concerns the ED cation disorder. It has been reported that for $CH_3NH_3PbI_3$ the conduction band minimum and the valence band maximum is mainly created by lead and iodine states and the contribution of the organic part does not affect the bands directly.[27] However, an interaction between the organic cation and the inorganic components of the structure plays an important role. Through the hydrogen bonding, the dynamic of the organic cation can change the band gap.[26] Therefore, similar to Motta et al. and Etienne et al., we assume that the organic cation position and mobility changes of the electron band structure and might be the origin for the increase of the carrier diffusion length and, consequently, better photoconductivity of the crystal with higher mobility of ED cation.[18,28]

**Thermal stability**

It has been reasoned that the thermal stability of organic–inorganic lead halide perovskites is similar as compared to the 3D perovskites containing lower chain amines, e.g. methylammonium or formamidinium cation.11 EDPbI4 shows very similar behaviour in TGA measurements (Fig. S3†) in comparison to CH3NH3PbCl3, CH3NH3PbBr3 and CH3NH3PbI3 powders (Fig. S4†). These organic–inorganic perovskites start to lose the organic components (NH3CH2CH2NH3I2, CH3NH3I, CH3NH3Br, CH3NH3Cl) under 573 K (the CH3NH3Cl under 473 K) and continues on at the isothermal 573 K.

**Stability of crystals against moisture**

The photovoltaic properties of the pristine crystal were studied once again ten months after it was synthesized. This crystal had been held at ambient conditions. At room temperature (293 K), the current density of this aged crystal decreased by 40% in comparison to its fresh state at the same temperature (Fig. 6b). The current density–voltage (J–V) curves were measured at 293 K in the dark and under UV light illumination (365 nm) for this aged crystal. Then the crystal was annealed for 10 minutes at 423 K. After 30 minutes of cooling, the measurements were performed again for this thermally treated aged crystal. Comparison of the measurements before and after heating of the aged crystal (Fig. 6c) shows almost identical increase of the current density induced by the thermal treatment. However, we can see that the aged crystal has degraded and its photoconductivity decreased by 40% in comparison to the freshly prepared pristine crystal (Fig. 6b). We suppose that the degradation is mainly associated with incorporation of water that is typical for other lead iodide perovskite related compounds.29

**CONCLUSION**

After thermal treatment we obtained ethylenediammonium lead iodide, NH3CH2CH2NH3PbI4, crystals possessing an arrangement of the PbI6 octahedra identical to the pristine crystals, but increased ethylenediammonium cation disorder. It is important to notice that the modification of the mutual positioning between the inorganic framework and ethylenediammonium cation might allow the synthesis of different dimensions of crystal structures (for instance, 1D and 2D) in the future. We observed a link between the increased disorder of ethylenediammonium in the thermally treated crystal and its photoconductivity. Measurements of the current–voltage curves revealed that after the thermal treatment the photoconductivity of the material improves by 40%. These results highlight the fragility and delicate structural– property of NH3CH2CH2NH3PbI4 to thermal treatment, which might be a general characteristic to all organic–inorganic low-dimensional halide perovskites-related compounds.

**ACKNOWLEDGMENTS**

The authors are grateful for the financial support of the European Research Council (No. 10306): grant PICOPROP No. 670918 and grant PICOPROP4CT No. 790341

**Table 1** Crystallographic data of NH3CH2CH2NH3PbI4

|  | Pristine crystal | Thermally treated crystal |
|---|---|---|
| Chemical formula | $I_4Pb \cdot C_2H_{10}N_2O_{0.042(4)}$ | $I_4Pb \cdot C_2H_{10}N_2O_{0.042(5)}$ |
| Formula weight | 777.60 | 777.60 |
| Space group | Trigonal, $R\bar{3}$ | Trigonal, $R\bar{3}$ |
| Temperature (K) | 293 | 293 |
| $a, c$ (Å) | 14.5906 (1), 32.7775 (5) | 14.5645 (1), 32.7195 (4) |
| $V$ (Å$^3$) | 6043.00 (14) | 6010.75 (13) |
| $Z$ | 18 | 18 |
| Wavelength (Å) | 0.7129 | 0.7129 |
| Crystal size (mm) | 0.03 × 0.02 × 0.003 | 0.03 × 0.02 × 0.003 |
| $\mu$ (mm$^{-1}$) | 21.89 | 22.01 |
| $R_{int}$ (%) | 4.2 | 3.8 |
| Diffractometer | PILATUS@SNBL[20] | PILATUS@SNBL[20] |
| Absorption correction | Multi-scan | Multi-scan |
| No. of measured, independent and observed reflections | 21 957, 3078, 2372 | 12 312, 2998, 2450 |
| $(\sin\theta/\lambda)_{min}$ (Å$^{-1}$) | 0.676 | 0.677 |
| $R[F^2 > 2\sigma(F^2)]$, $wR(F^2)$, $S$ | 0.032, 0.038, 1.49 | 0.040, 0.049, 1.89 |

**Table 2** Geometry characteristics of the ethylenediammonium cation

|  | Pristine | Thermally treated crystal | | |
|---|---|---|---|---|
|  |  | Dynamic model | Split model | Split C1 & C2 model |
| N1–C1 dist. (Å) | 1.432(15) | 1.51(2) | 1.41(4) | 1.48(3) & 1.49(3) |
| C1–C2 dist. (Å) | 1.449(13) | 1.29(2) | 1.43(5) | 1.40(7) & 1.50(5) |
| C2–N2 dist. (Å) | 1.508(16) | 1.43(2) | 1.52(4) | 1.60(7) & 1.39(7) |
| ΔC1 dist. (Å) | — | — | 0.90(9) | 0.86(3) |
| ΔC2 dist. (Å) | — | — | 0.51(9) | 0.77(3) |
| ΔN1 dist. (Å) | — | — | 0.25(7) | — |
| ΔN2 dist. (Å) | — | — | 0.34(9) | — |
| N1–C1–C2 angle (°) | 114.4(10) | 120.5(18) | 110(3) | 113(3) & 108(3) |
| C1–C2–N2 angle (°) | 111.3(9) | 119.3(17) | 116(3) | 108(4) & 117(3) |
| Deviation of C2 from the N1C1N2 cation plane (Å) | 0.13(3) | 0.03(5) | 0.15(9) | (i) N1C1aN2: C2a − 0.15(5), C2b − 0.45(4); (ii) N1C1bN2: C2a − 0.52(2), C2b − 0.18(3) |

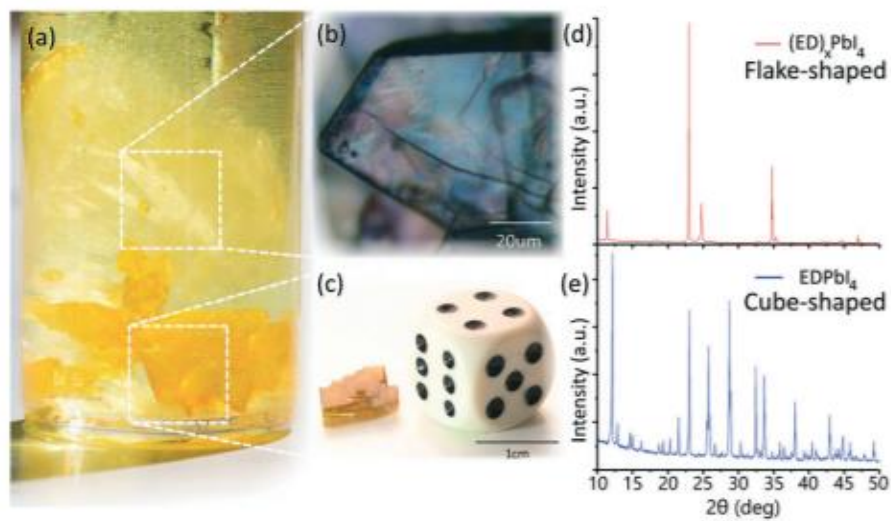

**Figure 1** NH3CH2CH2NH3PbI4 crystals. (a) Vial with two phases: (b) and (d) flake-shaped phase and (c) and (e) cube-shaped phase.

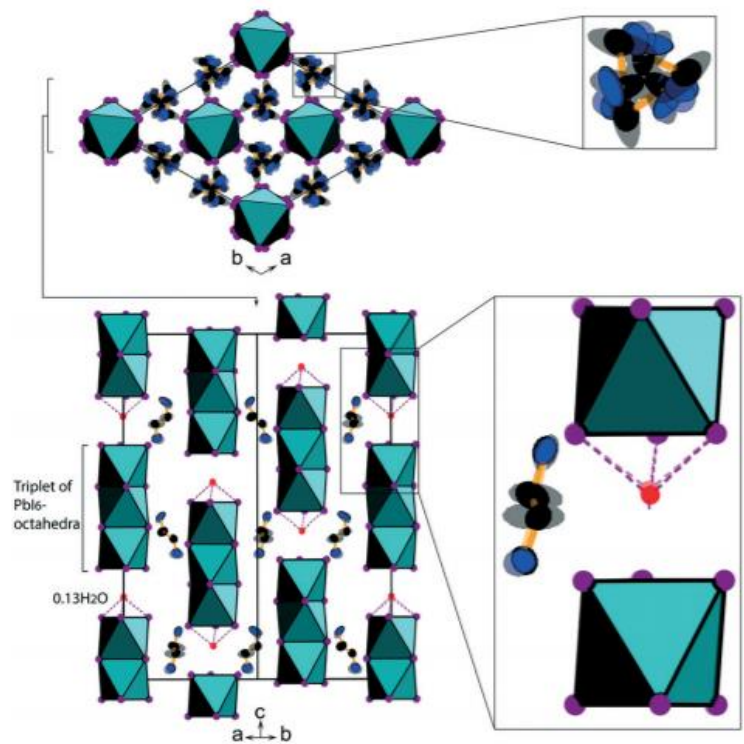

**Figure 2** Crystal structure of NH3CH2CH2NH3PbI4. Superposition of the pristine and thermally treated crystal structures is shown in two projections. The thermally treated crystal is drawn with the 50% opacity. Topology of the crystal structures is identical. The difference mainly concerns atomic displacement parameters of N and C in the organic cation, which is shown for the dynamic model of the C and N atoms disorder.

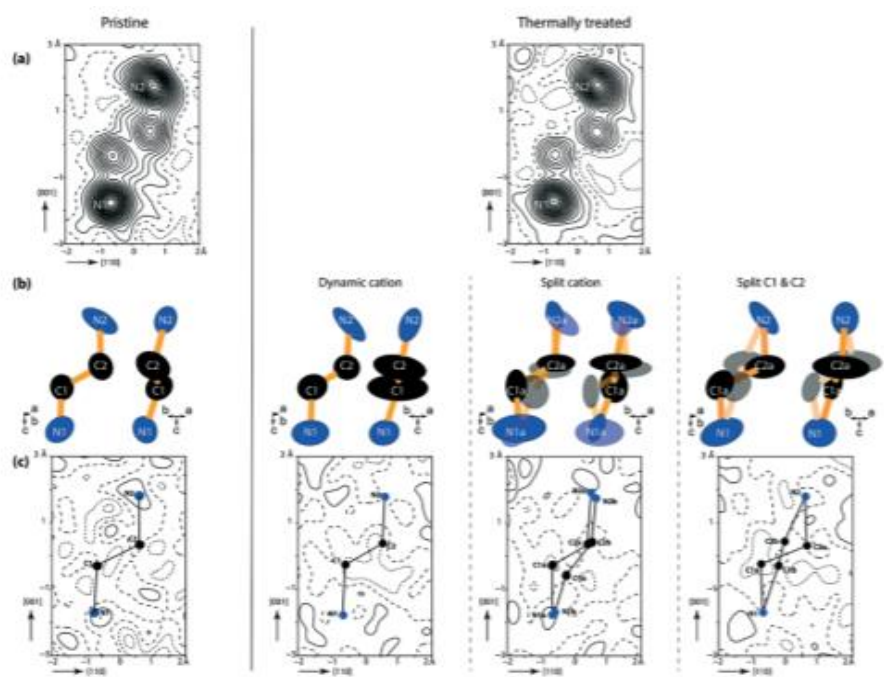

**Figure 3** Illustration of the ED organic cation localization with the difference electron density mapping in the pristine and thermally treated crystals. (a) The starting positions of the N and C atoms are determined from the maps calculated after the refinement of Pb and I atoms. (b) Two

projections of the cation obtained after refinement of Pb, I, N and C atoms. Unusually elongated ADP ellipsoids (50% probability) of N and C atoms cause considering of three models of the cation disorder in the thermally treated crystal. Split cations are shown with 100% and 50% opacity. (c) The maps obtained after refinement of Pb, I, N and C atoms. The residual electron density is indifferently low in all three models of the cation disordering in the thermally treated crystal. In (a) and (c), the solid dashed and striped lines correspond to positive, negative and zero contours, respectively, drawn with a step of 0.2 e Å$^{-3}$.

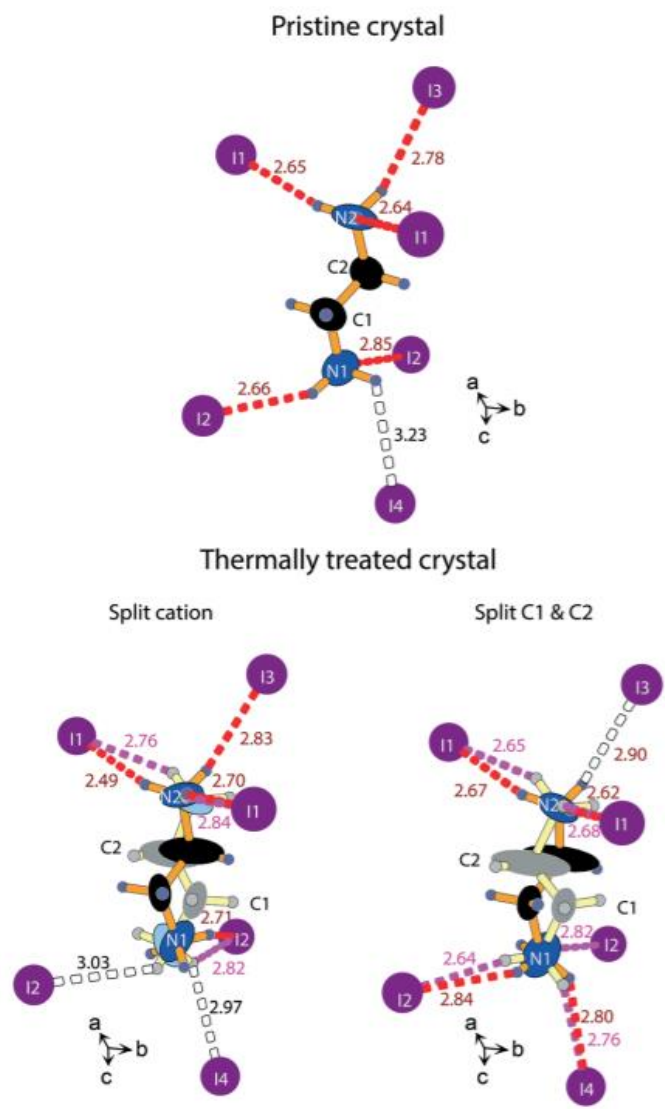

**Figure 4** Comparison of the N–H⋯I interactions in the pristine and two models of the thermally treated crystals of EDPbI4. The ED cation, H3N1–C1H2–C2H2–N2H3, and surrounding I atoms are shown in the identical perspective projection for each model. The distance N–H

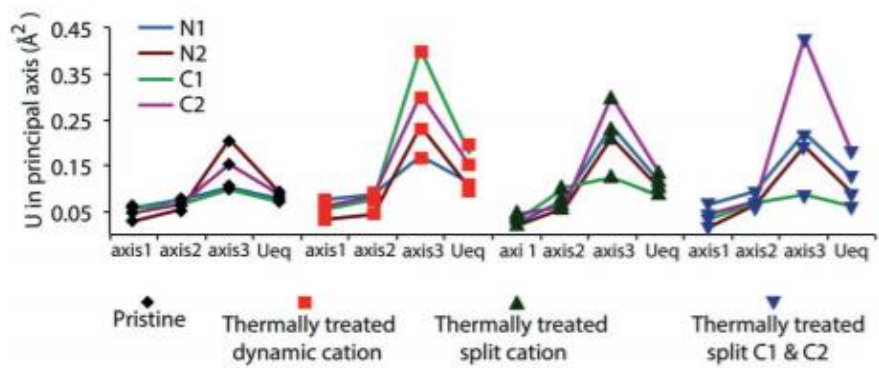

**Figure 5** Comparison of the displacement parameters of C and N atoms of ED cation in the pristine crystal and three models of the thermally treated crystal. The values of the principal axes of ADP ellipsoid and Ueq are displayed for each atom in each indicated model. The Ueq values are systematically lower in the pristine crystal.

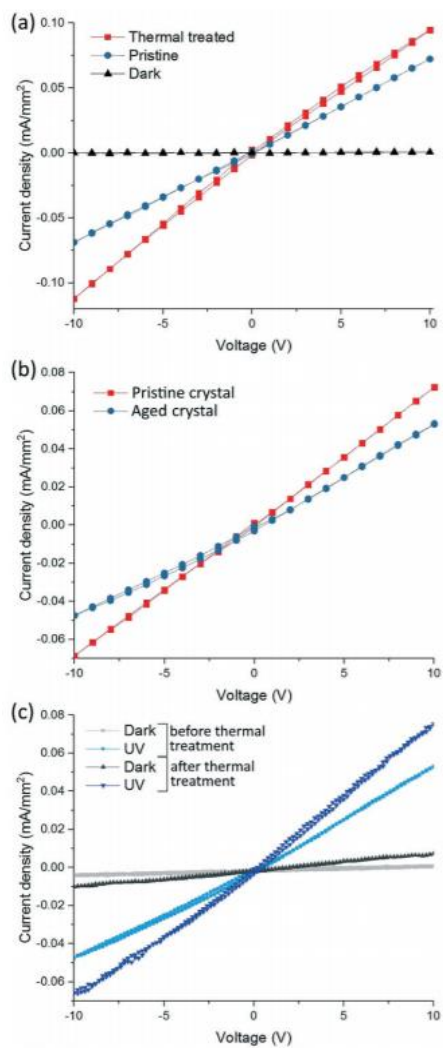

**Figure 6** The current density–voltage (J–V) curves of NH3CH2CH2NH3PbI4 crystal. (a) Comparison of the freshly prepared pristine and thermally treated crystals. (b) Comparison of the freshly prepared and aged pristine crystal. (c) Comparison of the aged crystal before and after thermal treatment, current density increased by 36%.